\documentclass[reprint,superscriptaddress,amsmath,amssymb,aps,pre]{revtex4-2}
\usepackage{graphicx}
\usepackage{dcolumn}
\usepackage{bm}
\usepackage{subfigure}
\usepackage{amsmath}
\usepackage{mathrsfs}
\usepackage{amsfonts}
\usepackage{color}
\usepackage{array} %
\usepackage{booktabs} %
\usepackage{rotating}
\usepackage[english]{babel}
\usepackage{amsthm,amsmath,amssymb}
\usepackage{float}
\usepackage[hidelinks]{hyperref}
\hypersetup{
    colorlinks=true,
    linkcolor=blue,
    filecolor=magenta,
    urlcolor=blue,
    citecolor=magenta,
    }   
\usepackage{mathptmx}

\DeclareMathOperator{\sech}{sech}
\usepackage[
text={7in,10in},centering,
height=10in,a4paper,hmargin={1.5cm,1.5cm},
]{geometry}

\begin{document}

\preprint{APS/123-QED}

\title{Magnetic lump motion in saturated ferromagnetic films}

\author{Xin-Wei Jin}
\affiliation{School of Physics, Northwest University, Xi'an 710127, China}
\affiliation{Department of Physics, Zhejiang Normal University, Jinhua 321004, China}%
\author{Shi-Jie Shen}
\affiliation{Department of Physics, Zhejiang Normal University, Jinhua 321004, China}%
\author{Zhan-Ying Yang}
\affiliation{School of Physics, Northwest University, Xi'an 710127, China}
\affiliation{Peng Huanwu Center for Fundamental Theory, Xi'an 710127, China}
\author{Ji Lin}%
\email{Corresponding author: linji@zjnu.edu.cn}
\affiliation{Department of Physics, Zhejiang Normal University, Jinhua 321004, China}%

\date{\today}

\begin{abstract}
In this paper, we study in detail the nonlinear propagation of magnetic soliton in a ferromagnetic film.
The sample is magnetized to saturation by an external field perpendicular to film plane.
A new generalized (2+1)-dimensional short-wave asymptotic model is derived.
The bilinear-like forms of this equation are constructed and exact magnetic line soliton solutions are exhibited.
It is observed that a series of stable lumps can be generated by an unstable magnetic soliton under Gaussian disturbance.
Such magnetic lumps are highly stable and can maintain their shapes and velocities during evolution or collision.
The interaction between lump and magnetic soliton, as well as interaction between two lumps, are numerically investigated.
We further discuss the nonlinear motion of lumps in ferrites with Gilbert-damping and inhomogeneous exchange effects.
The results show that the Gilbert-damping effects make the amplitude and velocity of the magnetic lump decay exponentially  during propagation.
And the shock waves are generated from a lump when quenching the strength of inhomogeneous exchange.
\end{abstract}

\maketitle


\section{Introduction}

The propagation of electromagnetic wave in ordered magnetic materials, especially in a ferromagnetic medium, plays a vital role in faster and higher density storage fields \cite{daniel1997soliton,veerakumar1998electromagnetic,hoefer2012propagation}.
In particular, magnetic soliton(MS), which exists in both ferro- and antiferro-magnets, is becoming a very promising information carrier because of its particle-like behavior and maneuverability \cite{kosevich1990magnetic,li2007soliton,kavitha2011breatherlike,tan2019propagation,fujimoto2019flemish,chai2020magnetic}. In the past few decades, a wide range of soliton-type propagation phenomena has been theoretically predicted \cite{leblond2007single,yu2021dark,sathishkumar2017oscillating,iacocca2014confined}, and some of them have been confirmed experimentally \cite{chai2021magnetic,mito2018geometrical}.

Indeed, wave propagation in ferromagnetic media is well-known as a highly nonlinear problem.
A complete description of all types of nonlinear excitations is governed by the Maxwell equations coupled with Landau-Lifschitz equation.
For this moment, let us notice that a fully nonlinear theory has not been developed. But the linear theory for sufficiently small amplitudes was established and validated experimentally \cite{walker1985properties}.
In order to obtain results valid in nonlinear regimes, or at least weakly nonlinear, one has to resort to intermediate models (by introducing a small perturbative parameter related to the soliton wavelength) \cite{leblond2008electromagnetic}.
These models include long-wave model \cite{leblond1994focusing,leblond2003new,nakata1991weak}, modulational asymptotic model \cite{leblond1999electromagnetic}, and short-wave model \cite{nguepjouo2014soliton,kuetche2016inhomogeneous,li2018rich,tchokouansi2016propagation}.
Both long-wave model and modulational asymptotic model are mainly used to explain and predict the behavior of large-scale phenomena owing to their long-wave-type approximate condition \cite{manna2006transverse}.
However, this condition is not always applicable because the scale of magnetic materials and devices are getting more refined and more sophisticated.  Moreover, the main practical interest of ferrites is that they propagate microwaves \cite{tchidjo2019dynamics,zhao2011abnormal}.
On the contrary, from the viewpoint of applied physics, the short-wave-type approximation is much more relevant to available experiments than the former one.

Since Kraenkel et al. first proposed the short-wave model \cite{kraenkel2000nonlinear}, quite a few related nonlinear evolution equations have been derived, which belong to the Kraenkel-Manna-Merle (KMM) system \cite{nguepjouo2014soliton,kuetche2016inhomogeneous,leblond2008nonlinear,leblond2009short,nguepjouo2019inhomogeneous}.
Some significant works have been devoted to searching and explaining different excitation patterns of ferromagnetic insulators.
As for (1+1)-dimensional KMM system, the existence of multi-valued waveguide channel solutions has been verified, and the nonlinear interaction properties were investigated between the localized waves alongside the depiction of their energy densities \cite{nguepjouo2014soliton}.
By applying the Hirota bilinear transformation method, the one- and two-soliton solutions were constructed while studying in details the solitons scattering properties \cite{kuetche2016inhomogeneous}. This system is also solvable using the inverse scattering method \cite{tchokouansi2016propagation}.
It is noteworthy that this system possesses the loop-soliton and spike-like soliton \cite{jin2020contributions,kuetche2015investigation},
and the magnetic loop-soliton dynamics have been extensively studied \cite{saravanan2018engineering,jin2020rogue,tchokouansi2019traveling}.
The propagation of electromagnetic waves in higher-dimensional ideal ferromagnets has also been studied, corresponding to the (2+1)-dimensional KMM system \cite{manna2006transverse,leblond2009short,seadawy2020arising,kuetche2011fractal}. The analytical one-line-soliton solution as well as its transverse stability have been reported \cite{manna2006transverse}. It has been shown that these structures were stable under certain conditions.

On the other hand, most previous studies have only focused on the propagation of MS in ideal ferrites, which means some important properties of the magnetic material were neglected. The main reason is that the nonlinear wave equation describing the propagation of electromagnetic waves in non-ideal ferromagnetic materials is no longer integrable. However, the Gilbert-damping and inhomogeneous exchange effects are essential features in a real ferromagnetic film, and their connection with MS motion is an important issue that has not been explored so far.
In this paper, we aim to investigate theoretically and numerically the dynamics of the MS in a ferromagnetic film including damping and the inhomogeneous exchange effect.
The rest of this paper is organized as follows.
In Section 2, we review the physical background and derive a new (2+1)-dimensional short-wave asymptotic model in ferromagnetic media.
In Section 3, the bilinear-like form of the reduced system is constructed and the analytical MS solutions are acquired.
In Section 4, the transmission stability of the magnetic soliton is numerically explored. The results show that \textcolor{red}{an} unstable MS will split to some magnetic lumps by a small perturbation. The motions of these lumps under the influence of damping and inhomogeneous exchange are analysed in detail.
We end this work in Section 5 with a brief conclusion and perspectives.
\section{Physical Background}
\subsection{Basic equations}
The physical system under consideration is a saturated magnetized ferrite film lying in the $x-y$ plane, as shown in Fig. 1. Different from Ref. \cite{nguepjouo2019inhomogeneous}, we consider the external field $\textbf{H}_{0}^{\infty}$ perpendicular to the film, i.e., $\bf{M}$$_{0}=(0,0,m)$.
So the transverse drift is avoided.
The typical thickness of the film is about 0.5mm, and the width is approximately 10mm.
We assume the propagation distance is large enough with regard to the wavelength, say more than 50cm.
The evolution of the magnetic field $\textbf{H}$ and the magnetization density $\textbf{M}$ is governed by the Maxwell equations coupled with Landau-Lifschitz-Gilbert equation, which read as
\begin{subequations}
\begin{align}
&-\nabla(\nabla\cdot\textbf{H})+\Delta\textbf{H}=\frac{1}{c^{2}}\frac{\partial^{2}}{\partial t^{2}}(\textbf{H}+\textbf{M}),\\
&\frac{\partial}{\partial t}\textbf{M}=-\gamma\mu_{0}\textbf{M}\times\textbf{H}_{\rm{eff}}+\frac{\sigma}{M_{s}}\textbf{M}\times\frac{\partial}{\partial t}\textbf{M},
\end{align}
\end{subequations}
where $c=1/\sqrt{\mu_{0}\tilde{\epsilon}}$ is the speed of light with the scalar permittivity $\tilde{\epsilon}$ of the medium, $\gamma$ is the gyromagnetic ratio, $\mu_{0}$ being the magnetic permeability of the vacuum, $\sigma$ is the damping constant, and $M_{s}$ is the saturation magnetization.
The effective field $\textbf{H}_{{\rm eff}}$ is given by \cite{leblond2008nonlinear}
\begin{equation}
\begin{split}
\textbf{H}_{\rm{eff}}=\textbf{H}-\beta\textbf{n}(\textbf{n}\cdot\textbf{M})+\alpha\Delta\textbf{M}.
\end{split}
\end{equation}
\begin{figure}[htbp]
\centering
\vspace{-0.2cm} %
\includegraphics[width=6.5cm]{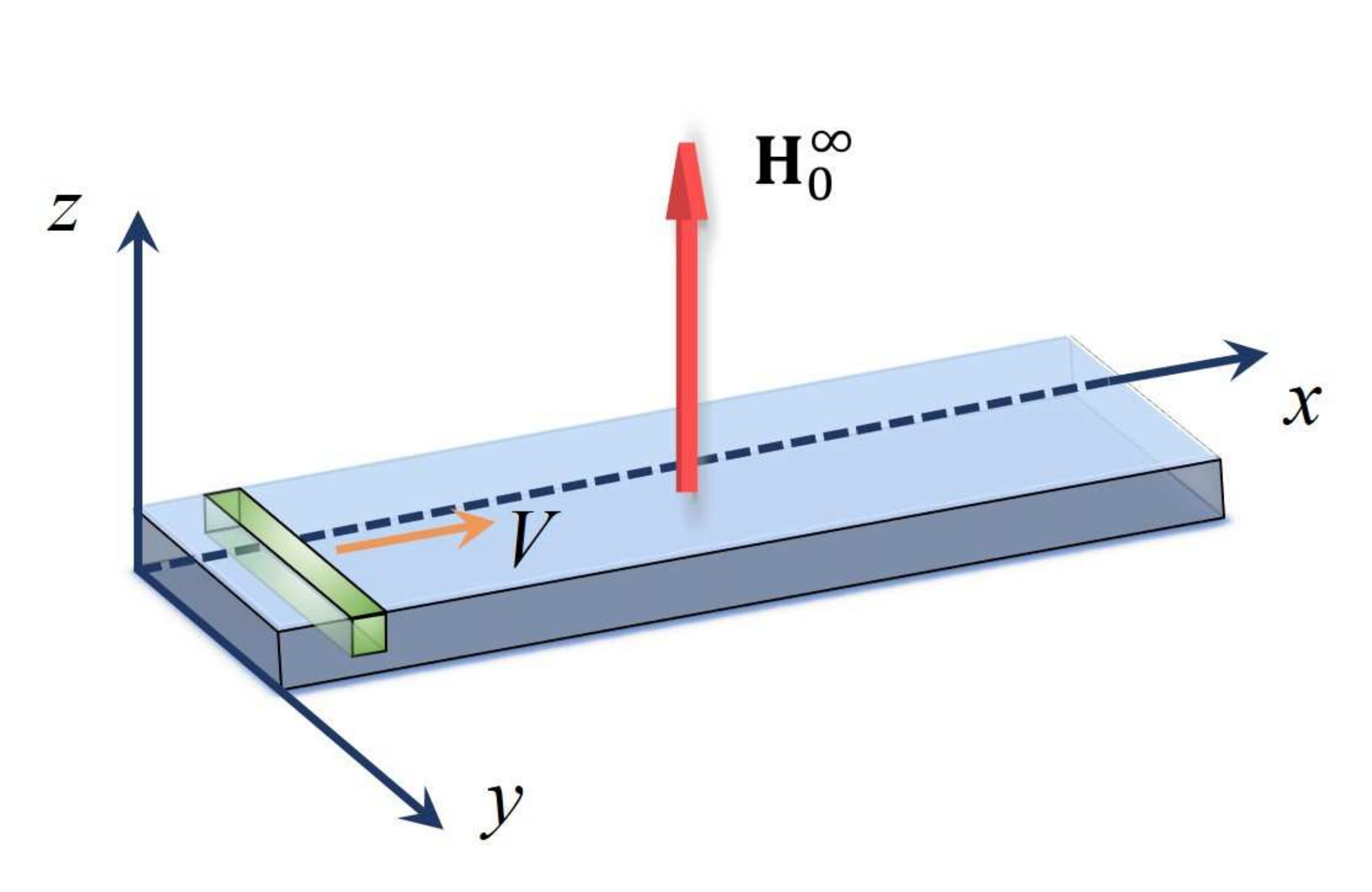}
\caption{Ferrite film under consideration. The sample is magnetized to saturation by long strong magnetic field $\textbf{H}_{0}^{\infty}$ applied in the $z$-direction. The $x$-direction of the short wave propagation is perpendicular to the direction of static magnetization.}
\end{figure}
Here $\alpha$ and $\beta$ are the constants of the inhomogeneous exchange and the magnet anisotropy ($\beta>0$ corresponds to the easy-plane case), respectively.
For a simple tractability, the unit vector \textbf{n} of the anisotropy axis is assumed to be along the $z$ axis (i.e., $\textbf{n}\equiv \textbf{e}_{z}$).
In order to transform the above systems to dimensionless equation, we rescale the quantities $\textbf{M}$, $\textbf{H}$, and $t$ into
$\mu_{0}\gamma\textbf{M}/c$, $\mu_{0}\gamma\textbf{H}/c$, and $ct$.
Thus, the constants $\mu_{0}\gamma/c$ and $c$ in Eqs.(2) and (3) are replaced by 1, $M_{s}$ by $m=\mu_{0}\gamma M_{s}/c$, and $\sigma$ by $\tilde{\sigma}=\sigma/\mu_{0}\gamma$ \cite{leblond2008nonlinear}.
\subsection{Linear analysis}
To study the linear regime we look at a small perturbation of a given solution. Equations (1) are linearized about the steady state:
\begin{equation}
\begin{split}
\textbf{M}_{0}=(0 ,0 ,m),\ \ \textbf{H}_{0}=\mu\textbf{M}_{0}.
\end{split}
\end{equation}
where $\mu$ is the strength of the internal magnetic field.
Before proceeding further we assume that the ferromagnetic materials have weak damping $\bar{\sigma}\sim\epsilon\tilde{\sigma}$.
The exchange interaction parameter $\alpha$ and anisotropy parameter $\beta$ are of order $\epsilon^{2}$ and $\epsilon^{3}$, respectively (i.e. $\bar{\alpha}=\epsilon^2\alpha, \bar{\beta}=\epsilon^3\beta$).
Let us seek for the plane wave perturbation solution propagating along the $x$-direction such as
\begin{equation}
\begin{split}
&\textbf{M}=\textbf{M}_{0}+\epsilon\textbf{m}\exp[i(kx+ly-\omega t)],\\
&\textbf{H}=\textbf{H}_{0}+\epsilon\textbf{h}\exp[i(kx+ly-\omega t)],
\end{split}
\end{equation}
where $k$ and $l$ are the wave numbers in the $x$ and $y$ directions, $\omega$ is the frequency.
Vectors $\textbf{m}=(m^{x},m^{y},m^{z})$ and $\textbf{h}=(h^{x},h^{y},h^{z})$ are arbitrary real scalar quantities.

Substituting Eq. (4) into (1) and (2) in the linear limit, it is reduced to
$$
\begin{pmatrix}
\omega^{2} & 0 & 0 & \omega^{2}-l^{2} & kl & 0\\
0 & \omega^{2} & 0 & kl & \omega^{2}-k^{2} & 0\\
0 & 0 & \omega^{2} & 0 & 0 & \omega^{2}-k^{2}-l^{2}\\
-i\omega & m\mu & 0 & 0 & -m & 0\\
-m\mu & -i\omega & 0 & m & 0 & 0\\
0 & 0 & -i\omega & 0 & 0 & 0
\end{pmatrix}
\cdot
\begin{pmatrix}
m_{x}\\ m_{y}\\ m_{z}\\ h_{x}\\ h_{y}\\ h_{z}
\end{pmatrix}
=0
$$
Then we obtain the following dispersion relation
\begin{equation}
m^{2}(\mu+1)\left[\mu(k^{2}+l^{2}-\omega^{2})-\omega^{2}\right]-\omega^{2}(k^{2}+l^{2}-\omega^{2})=0
\end{equation}
Note that we focus on studying the short-wave approximation $k\to\infty$ [2]. It comes $k_{0}\sim\epsilon^{-1}$ through a small parameter $\epsilon\ll1$ linked to the magnitude of the wavelength.
Consequently, the frequency expands accordingly as
\begin{equation}
\begin{split}
\omega=\omega_{-1}\epsilon^{-1}+\omega_{1}\epsilon+\omega_{3}\epsilon^{3}+....
\end{split}
\end{equation}
This assumption guarantees the phase velocity $\omega(k)/k$ and the group velocity $\partial\omega/\partial k$ are always bounded [3].
Now, replacing Eq. (6) into the dispersion relation above, we obtain a set of equations:\\
$\bullet$\quad At order of $\epsilon^{-4}$: $\omega_{-1}=\pm k_{0}$\\
$\bullet$\quad At order of $\epsilon^{-2}$: $\omega_{1}=\left[(\mu+1)m^{2}+l^{2}\right]/2k_{0}$\\
$\bullet$\quad higher order equations which determines $\omega_{3}$, $\omega_{5}$,...\par
The direction of the wave propagation is assumed to be close to the $x$ axis, thus $y$ variable gives only account of a slow transverse deviation\cite{leblond2008reductive,leblond2009two}. Therefore $l$ is assumed to be very small with respect to $k$ and we write $l=l_{0}$ of order 0 with respect to $\epsilon$. The phase up to order $\epsilon$ is thus
$(x-t)/\epsilon+l_{0}y-\epsilon\omega_{1}t,$
which motivates the introduction of new variables:
\begin{equation}
\zeta=\frac{1}{\epsilon}(x-Vt),\ \ y=y,\ \ \tau=\epsilon t.
\end{equation}
The variable $\zeta$ describes the shape of the wave propagating at speed $V$; it assumes a short wavelength about $1/\epsilon$. The slow time variable $\tau$ accounts for the propagation during very long time on very large distances with regard to the wavelength. The transverse variable $y$ has an intermediate scale, as in KP-type expansions \cite{leblond2009two,manna2006transverse}
\subsection{Multiple scale approach}
In order to derive the nonlinear model, fields \textbf{M} and \textbf{H} are expanded in power series of $\epsilon$ as
\begin{equation}
\begin{split}
&\textbf{M}=\textbf{M}_{0}+\epsilon\textbf{M}_{1}+\epsilon^{2}\textbf{M}_{2}+\epsilon^{3}\textbf{M}_{3}+...,\\
&\textbf{H}=\textbf{H}_{0}+\epsilon\textbf{H}_{1}+\epsilon^{2}\textbf{H}_{2}+\epsilon^{3}\textbf{H}_{3}+... .
\end{split}
\end{equation}
where $\textbf{M}_{0}$, $\textbf{H}_{0}$, $\textbf{M}_{1}$, $\textbf{H}_{1}$,...are functions of $(\zeta,y,\tau)$.
We consider the boundary conditions: $\underset{\zeta\to-\infty}{\lim}\textbf{M}_{0}=(0,0,m),\underset{\zeta\to-\infty}{\lim}\textbf{M}_{j}=\underset{\zeta\to-\infty}{\lim}\textbf{H}_{j}=0,(j\ne0)$.
We derive the following expressions by substituting Expansions (8) into equation (1):\\
$\bullet$\quad At order $\epsilon^{-2}$:\par
 \textbf{M}$_{0}$ is a constant vector \textbf{M}$_{0}$=(0,0,m),\\
$\bullet$\quad At order $\epsilon^{-1}$:\par
 $H_{0}^{x}=0,\ \  M_{1}^{y}=0,\ \  M_{1}^{z}=0$,\\
$\bullet$\quad At order $\epsilon^{0}$:\par
 $M_{1\zeta}^{x}=mH_{0}^{y}$,\par
 $M_{2\zeta\zeta}^{x}=-H_{2\zeta\zeta}^{x}-H_{1\zeta \tau}^{y}$\par
 $M_{2\zeta\zeta}^{y}=-H_{1\zeta y}^{x}+H_{0\zeta y}^{x}$\par
 $M_{2\zeta\zeta}^{z}=H_{2\zeta\tau}^{z}+H_{y y}^{z}$\\
$\bullet$\quad At order $\epsilon^{1}$:\par
 $M_{2\zeta}^{x}=-m H_{1}^{y}$\par
 $M_{2\zeta}^{y}=m\bar{\alpha}M_{1\zeta\zeta}^{x}+\bar{\sigma}M_{1\zeta^{x}}-M_{1}^{x}H_{0}^{z}+mH_{1}^{x}$\par
 $M_{2\zeta}^{z}=M_{1}^{x}H_{0}^{y}$\\
let us introduce some independent variables $X$ and $T$ defined as
$X=-m\zeta/2, Y=my, T=m\tau$.

After eliminating $\textbf{H}_{2}$ and $\textbf{M}_{2}$, we finally obtain the (2+1)-dimensional KMM equation:
\begin{equation}
\begin{split}
&C_{XT}=-BB_{X}+C_{YY},\\
&B_{XT}=BC_{X}+B_{YY}-sB_{X}+\rho B_{XX},
\end{split}
\end{equation}
where observables $B$, $C$ and constants $s$, $\rho$ are defined by
\begin{equation}
\begin{split}
&C=-X-\int^{X}(H^{z}_{0}/m)dX,\ \ B=M^{x}_{1}/2m,\\
&s=-\bar{\sigma}/2,\ \ \rho=\bar{\alpha}m^2/4.
\end{split}
\end{equation}
This equation is new, which describes the evolution of magnetization field $\textbf{M}$ and magnetic field $\textbf{H}$ within a ferrite film in presence of Gilbert-damping and inhomogeneous exchange.
The quantities $H_{0}$ and $M_{1}$ refer to the zeroth and first-order expansion coefficients of the external magnetic field and the magnetization, respectively.
For some simplicity, in the next, the independent variables $X$, $Y$ and $T$ will be rewritten as their lower cases $x$, $y$ and $t$, respectively.

\section{Hirota's bilinearization and soliton solutions of the (2+1)-dimensional KMM equation}
To explore soliton solutions for the (2+1)-dimensional KMM equation (9), we consider a specific dependent variable transformation
\begin{equation}
\begin{split}
B=\frac{G}{F},\ \ C=\delta x-2(\ln F)_{t}-2(\ln F)_{y},
\end{split}
\end{equation}
where $\delta$ is an arbitrary constant. Consequently, the bilinear-like forms of the (2+1)-dimensional KMM equation can be derived as follow
\begin{subequations}  \label{eq:2}
\begin{align}
&F\! \cdot\! (D_{x}D_{t}\! +\! sD_{x}\! -\! D_{y}^{2})G\! \cdot\! F\! +\! G\! \cdot\! (D_{x}D_{y}\! +\! D_{y}^{2})F\! \cdot\! F=\delta F^{2}G \label{eq:2A}\\
&\partial_{x}\! \left[\frac{G^{2}}{2F^{2}}\! -\! \frac{(D_{y}D_{t}\! +\! D_{t}^{2})F\! \cdot\!  F}{F^{2}}\right]\! +\! \partial_{y}\left[\frac{(D_{y}D_{t}\! +\! D_{t}^{2})F\! \cdot\! F}{F^{2}}\right]\! =\! 0 \label{eq:2B}
\end{align}
\end{subequations}
where $G$, $F$ are all differential functions of $(x,y,t)$ to be determined.
The symbols $D_{x}$, $D_{t}$ refer to the Hirota's operators with respect to the variable $x$, $t$, respectively.
In order to construct the solitary wave solutions of Eq.(6), we expand $G$ and $F$ with respect to a formal expansion parameter as
$G=\epsilon G_{1}+\epsilon^{3} G_{3}+\epsilon^{5} G_{5}+...,
F=1+\epsilon^{2}F_{2}+\epsilon^{4}F_{4}+\epsilon^{6} F_{6}+...,$
in which $\epsilon$ is a perturbation parameter and functions $G_{i}, F_{i},(i\!=\!1,2,3,...)$ are expansion coefficients of the above series. The one-soliton solution could be constructed by truncating the perturbation expansion of $G$ and $F$ as follow
\begin{equation}
G=e^{\eta_{1}}, \ \  F=1+\frac{k^{2}A^{2}}{16\delta^{2}}e^{2\eta_{1}}.
\end{equation}
Substituting these expressions into Eq.(9) and solving the bilinear system recursively, in the absence of damping, the analytical one-soliton solution of the (2+1)-dimensional KMM equation can be transformed into
\begin{equation}
B=\frac{2\delta}{k}\sech(\eta_{1}+\eta_{0}),\ C=\delta x-\frac{2\delta}{k}\left[\tanh(\eta_{1}\!+\!\eta_{0})\!+\!1\right],
\end{equation}
where $\eta_{1}\!=\!kx+ly+[(l^{2}-kl)/2k] t$, $\eta_{0}\!=\!\ln({k}/{4\delta})$, $k$ and $l$ are arbitrary real constants.
It should be noted that this soliton solution exists only when the damping is neglected $(s=0)$.
Similar to the procedure for constructing one-soliton solution, the two-soliton solution can be given by treating the truncated perturbation expansions of $G$ and $F$ as
\begin{subequations}
\begin{align}
&G\!=\!A_{1}e^{\xi_{1}}\!+\!A_{2}e^{\xi_{2}}\!+\!C_{12}e^{\xi_{1}\!+\!2\xi_{2}}\!+\!C_{21}e^{2\xi_{1}\!+\!\xi_{2}},\\
&F\!=\!1\!+\!B_{11}e^{2\xi_{1}}\!+\!B_{22}e^{2\xi_{2}}\!+\!B_{12}e^{\xi_{1}+\xi_{2}}\!+\!E_{12}e^{2\xi_{1}+2\xi_{2}},
\end{align}
\end{subequations}
where $A_{1},A_{2},k_{1},k_{2}$ are real constants, $\xi_{i}=k_{i}x+l_{i}y+\left[(l_{i}^{2}+\delta)/k_{i}\right]t,(i=1,2)$, and the remaining parameters have the following forms:
\begin{equation}
\begin{split}
&B_{ii}=\frac{A_{i}^{2}k_{i}^{2}}{16\delta^{2}},B_{12}=\frac{A_{1}A_{2}}{2\delta^{2}}\frac{k_{1}^{2}k_{2}^{2}}{k_{+}^{2}},k_{1}l_{2}=k_{2}l_{1},\\
&C_{ij}=\frac{A_{i}A_{j}^{2}}{16\delta^{2}}\frac{k_{j}^{2}k_{-}^{2}}{k_{+}^{2}},E_{12}=\frac{A_{1}^{2}A_{2}^{2}}{256\delta^4}\frac{k_{1}^{2}k_{2}^{2}k_{-}^{4}}{k_{+}^{4}},
\end{split}
\end{equation}
where $k_{+}=k_{1}+k_{2},\ k_{-}=k_{1}-k_{2}$.
Parameters $A_{i}$, $A_{j}$, $k_{i}$, $k_{j}$  and $l_{i},(i=1,2,j=3-i)$ are arbitrary real constants.
\begin{figure}[htbp]
\vspace{0cm} %
\subfigbottomskip=0pt %
\subfigure[]{
\begin{minipage}[t]{0.45\linewidth}
\centering
\includegraphics[width=4cm]{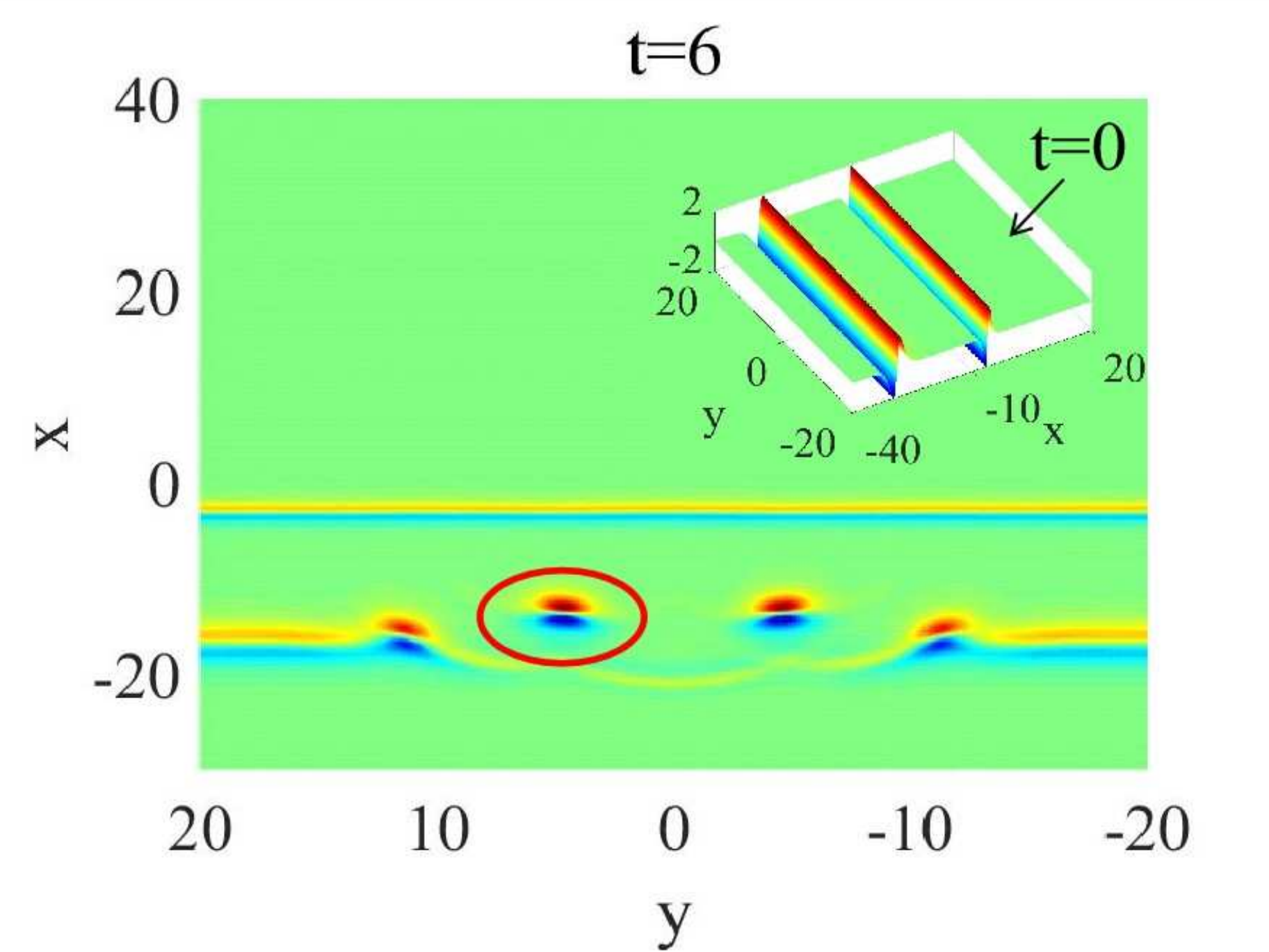}
\end{minipage}
}
\subfigure[]{
\begin{minipage}[t]{0.45\linewidth}
\centering
\includegraphics[width=4cm]{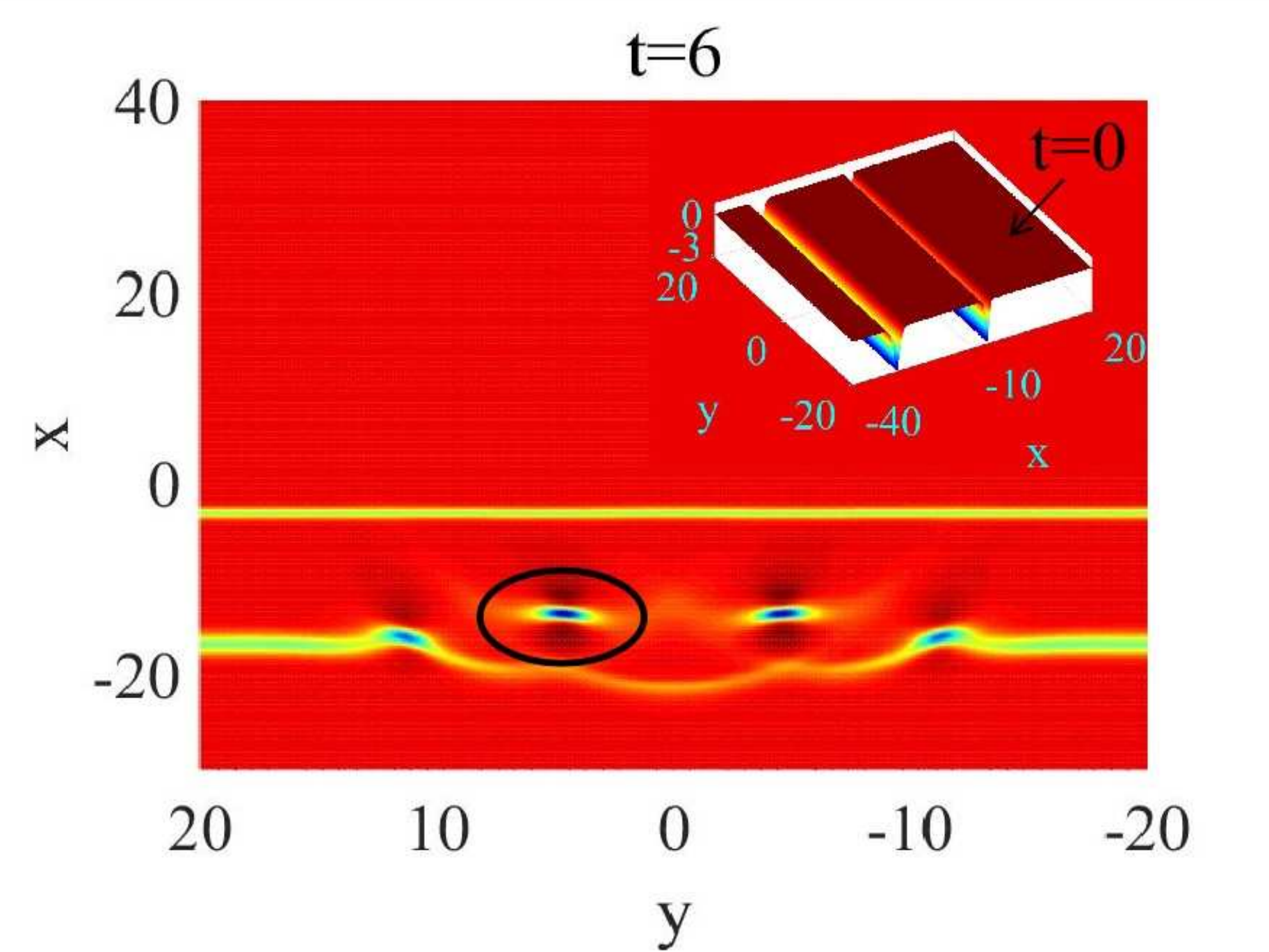}
\end{minipage}
}

\subfigure[]{
\begin{minipage}[t]{0.45\linewidth}
\centering
\includegraphics[width=4cm]{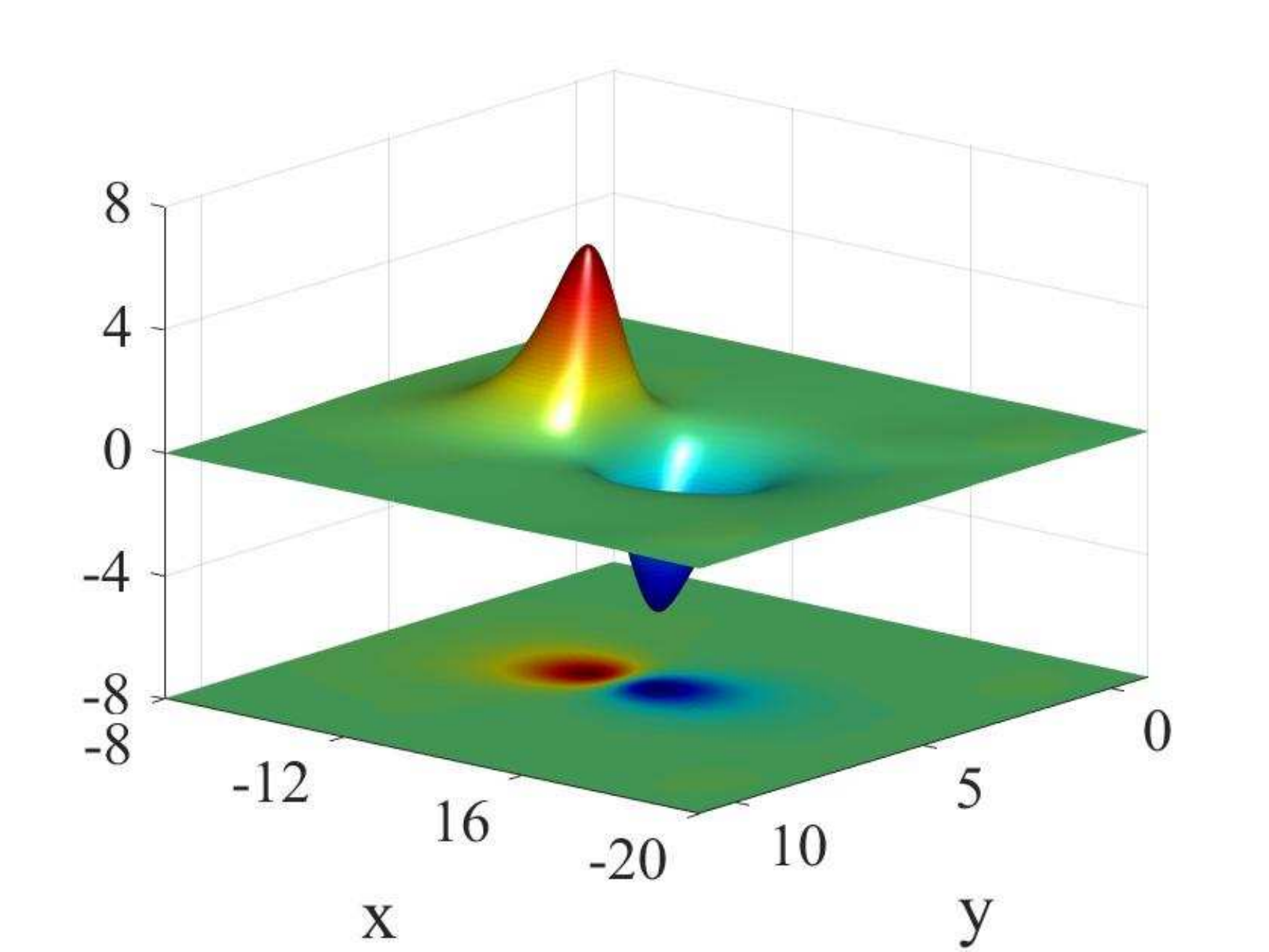}
\end{minipage}
}
\subfigure[]{
\begin{minipage}[t]{0.45\linewidth}
\centering
\includegraphics[width=4cm]{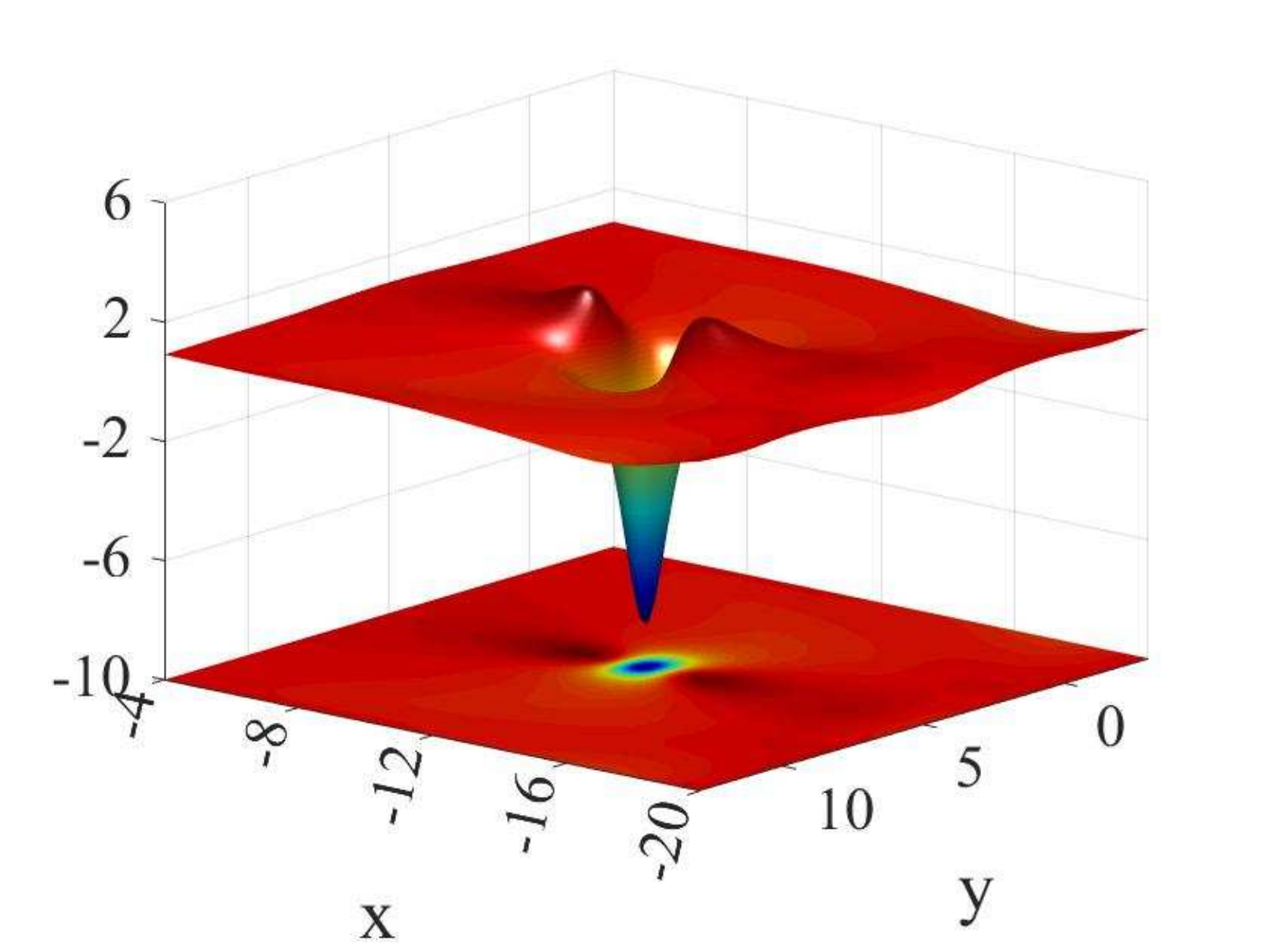}
\end{minipage}
}
\caption{Propagation of MS perturbed by a Gaussian disturbance. (a) Component $H^{z}$, (b) Component $H^{y}$, (c) and (d) are enlarged views of the indicated areas circled in red and black, respectively. The parameters are chosen as $A_{1}=A_{2}=1, \delta=-1, l_{1}=l_{2}=0, k_{1}=1, k_{2}=2, x_{0}=-29, b=0.1, x_{r}=1.5, y_{r}=2.5$ in (16) and (17).}
\end{figure}

\section{Numerical investigation of line-soliton and magnetic lumps}
\subsection{Unstable MS splits into lumps}

We now turn to the stability and interactions between MSs in a ferromagnetic film.
The initial data is a MS perturbed by some position-dependent Gaussian wave packets with the following expression:
\begin{equation}
\begin{split}
f=b\exp\left[-\left(\frac{x-x_{0}}{x_{r}}\right)^{2}-\left(\frac{y}{y_{r}}\right)^{2}\right],
\end{split}
\end{equation}
where $b,x_{r}$ and $y_{r}$ correspond to the shape of the wave packet and $x_{0}$ is related to the perturbation position.

The time evolution results clearly show the instability of the MS.
For small $b_{i}$, the MS will break up and eventually evolve into some stable two-dimensionally localized \emph{lumps}, as displayed in Figs. 2(a) and 2(b).
We observe that most of the energy is always propagated as a lump, even if its speed may differ from the input.
Such a magnetic lump is a solitary wave packet that maintains its shape and speed during propagation or collision.

A complete single lump of magnetic field component $H^{z}$ (component $H^{y}$) is circled in red (black) in Fig.2.
The enlarged views (see Figs.2(c) and 2(d)) provide a clear picture of the shape and contour map of the lump.
It can be found that component $H^{z}$ is a dipole-mode lump, whereas component $H^{y}$ is a standard KP-lump.
We also show the vector field of the magnetic lump in Fig.3(a).
Note that magnetic field component $H^{x}$ is zero, the magnetic field is always in the $y-z$ plane, hence the lump can be regarded as a $360^{\circ}$ domain wall localized in $x$ and $y$ directions. Fig.3(b) presents the magnetic field along $y\!=\!0$.  The blue and red arrows
correspond to the magnetic field intensity of component $H^{z}$, $H^{y}$, respectively.
The rest of this work is concerned with the propagation and interaction behavior of these lumps in ferrite medium.
\begin{figure}[htbp]
\vspace{-0cm}
\subfigbottomskip=0pt
\subfigure[]{
\begin{minipage}[t]{0.45\linewidth}
\centering
\includegraphics[width=4.8cm]{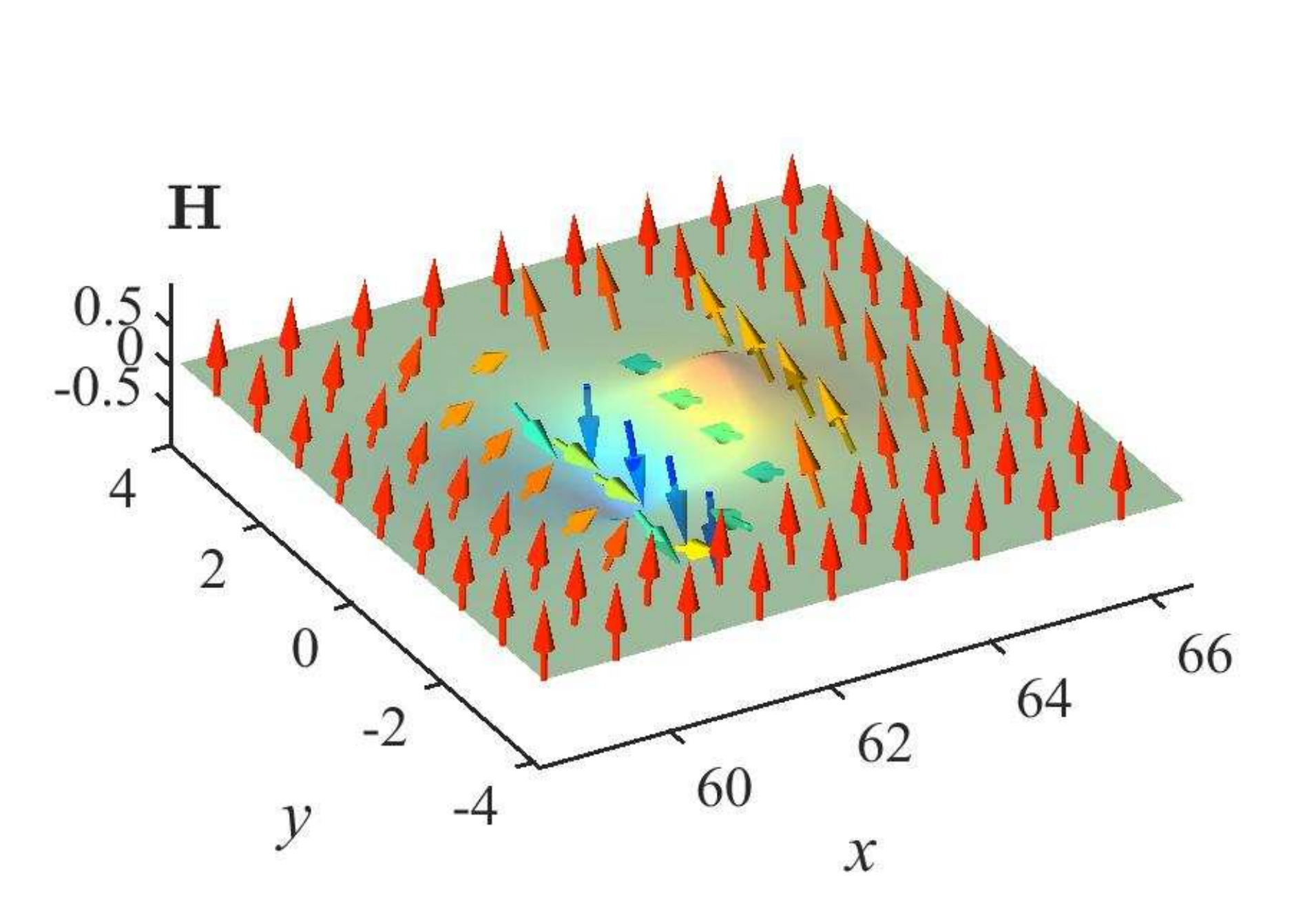}
\end{minipage}
}
\subfigure[]{
\begin{minipage}[t]{0.45\linewidth}
\centering
\includegraphics[width=3.8cm]{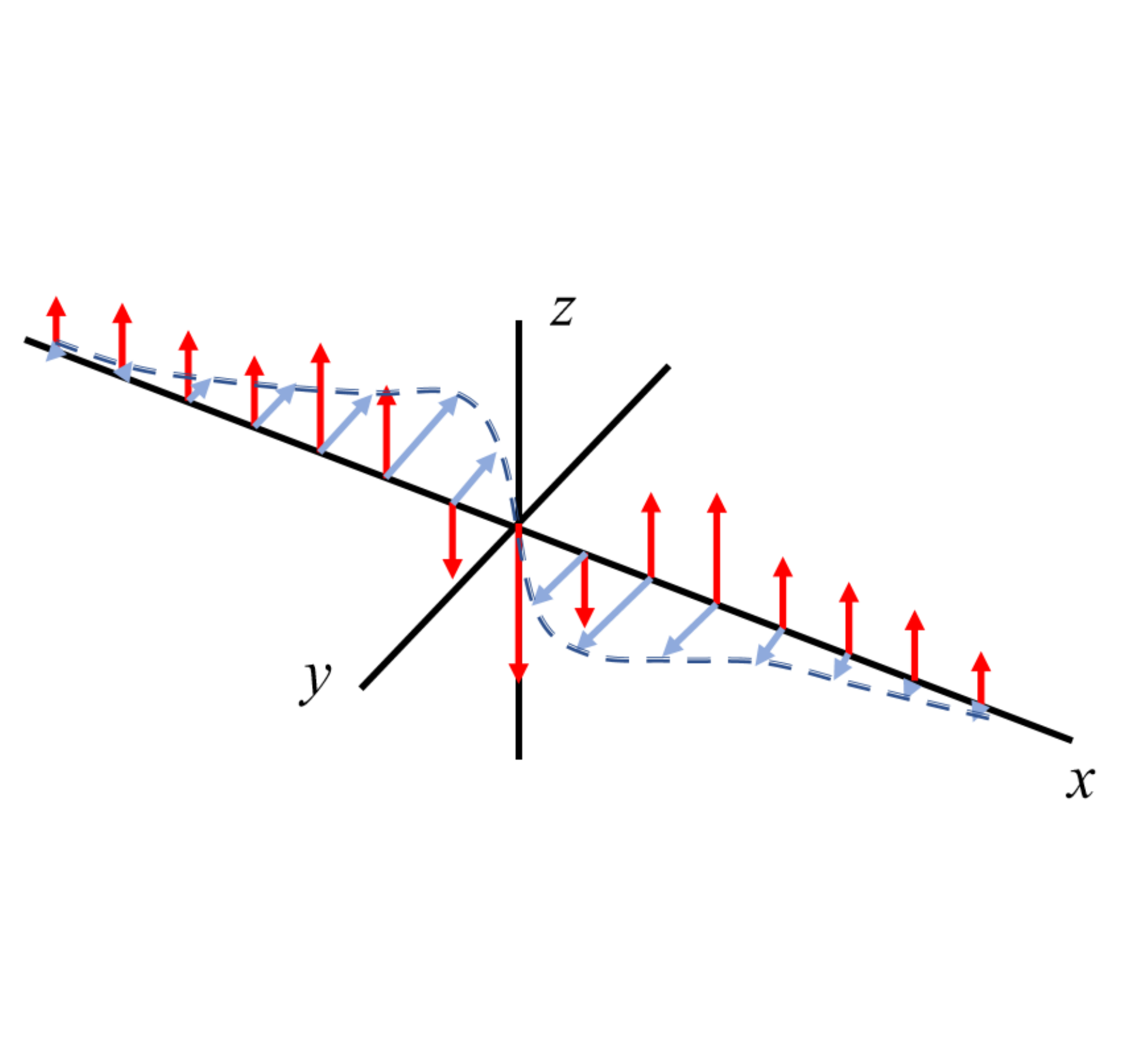}
\end{minipage}
}
\caption{(a) The vector field of the magnetic lump. (b) The magnetic lump along $y=0$. The blue and red arrows correspond to the magnetic field intensity of components $H^{z}$, $H^{y}$, respectively.}%
\end{figure}

\subsection{Lump motion in ferromagnets with damping or inhomogeneous exchange effects}
\begin{figure}[hp]
\centering
\includegraphics[width=6.2cm]{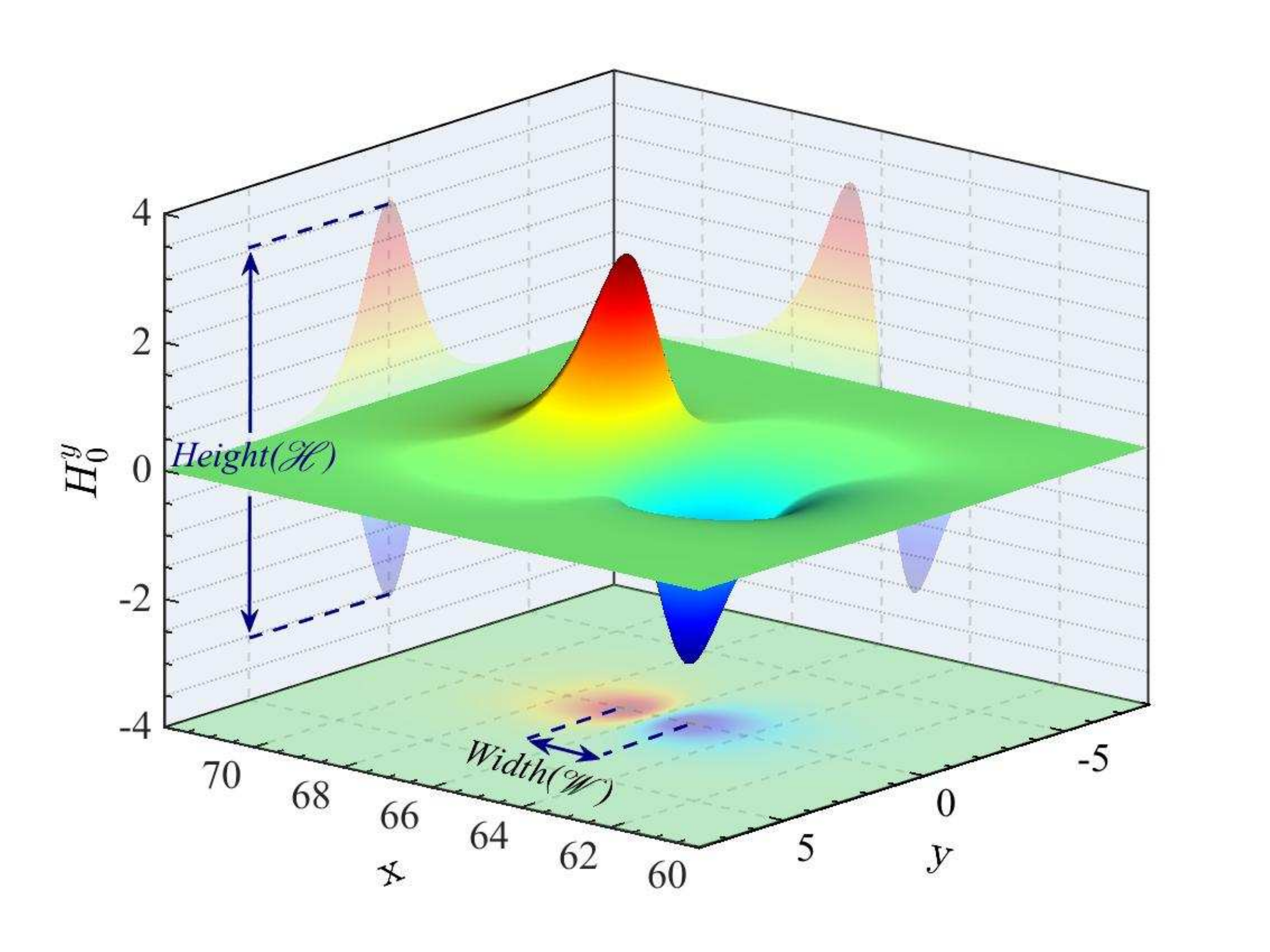}
\caption{Three dimensional projections of lump at $t=0$, $\mathcal{H}$ and $\mathcal{W}$ represent the definitions of lump height and width, respectively.}
\end{figure}

\begin{figure*}[htb]
\subfigure[]{
\begin{minipage}[t]{0.3\linewidth}
\centering
\includegraphics[width=5cm]{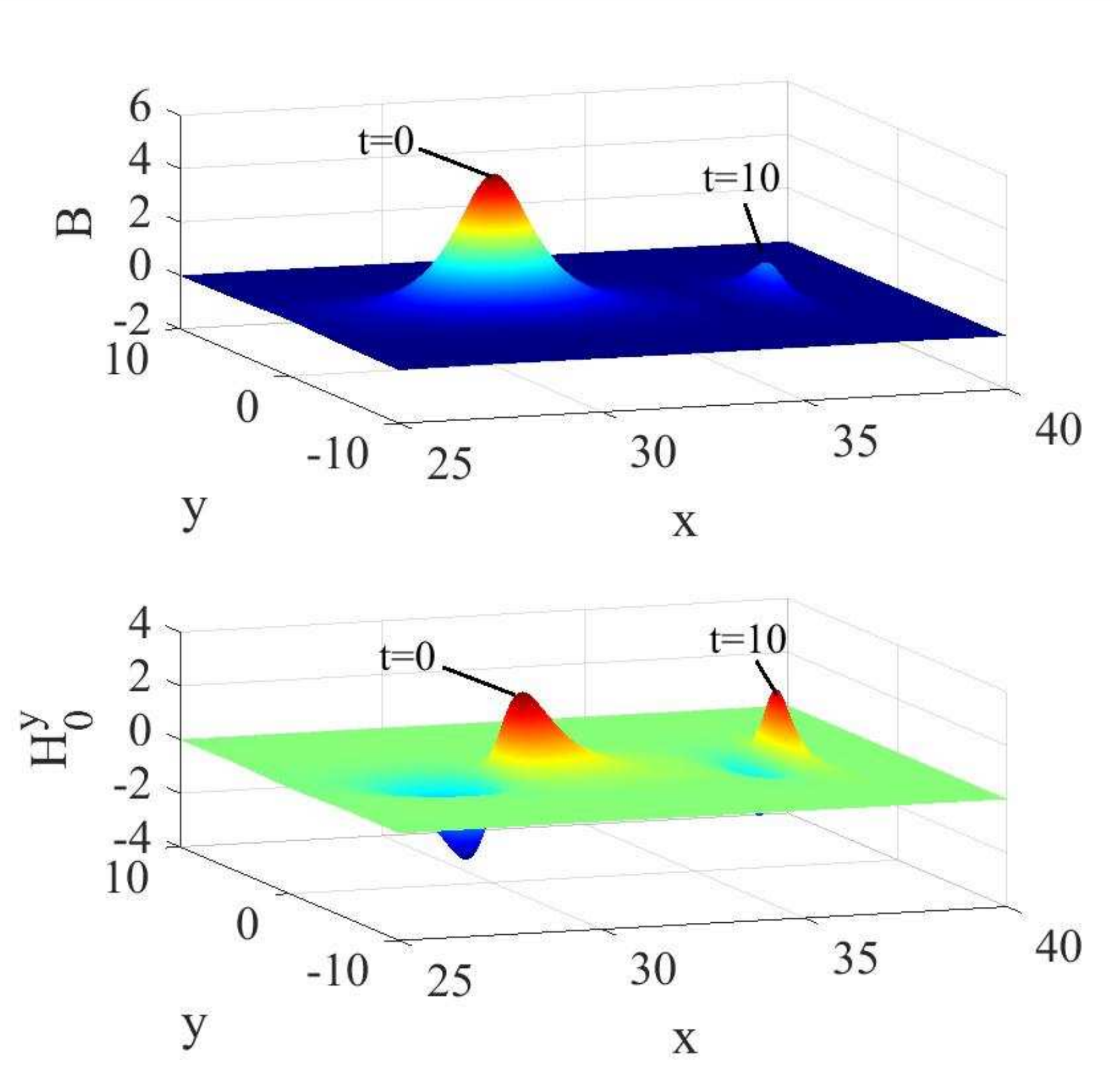}
\end{minipage}
}
\subfigure[]{
\begin{minipage}[t]{0.3\linewidth}
\centering
\includegraphics[width=5cm]{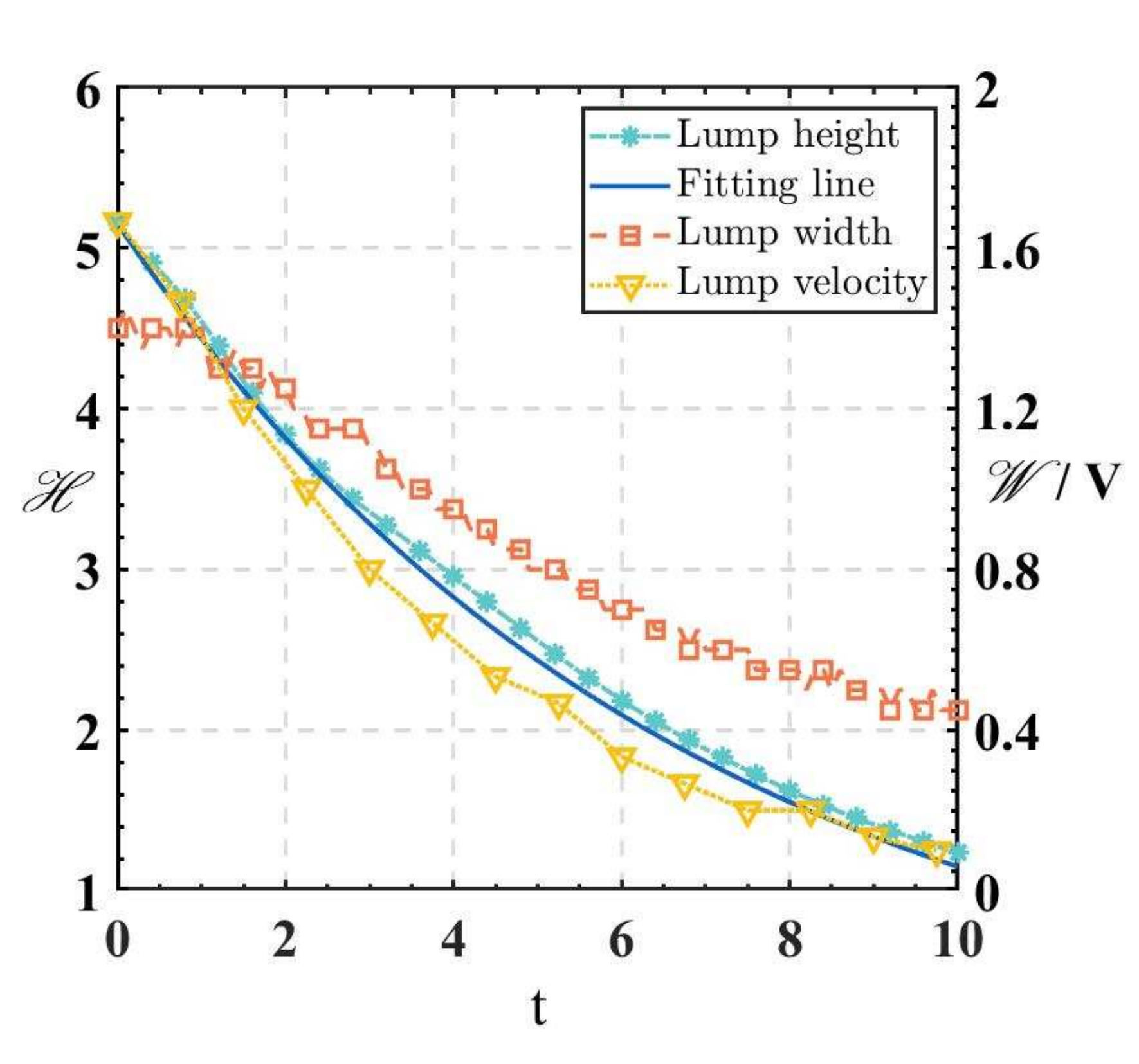}
\end{minipage}
}
\subfigure[]{
\begin{minipage}[t]{0.3\linewidth}
\centering
\includegraphics[width=5cm]{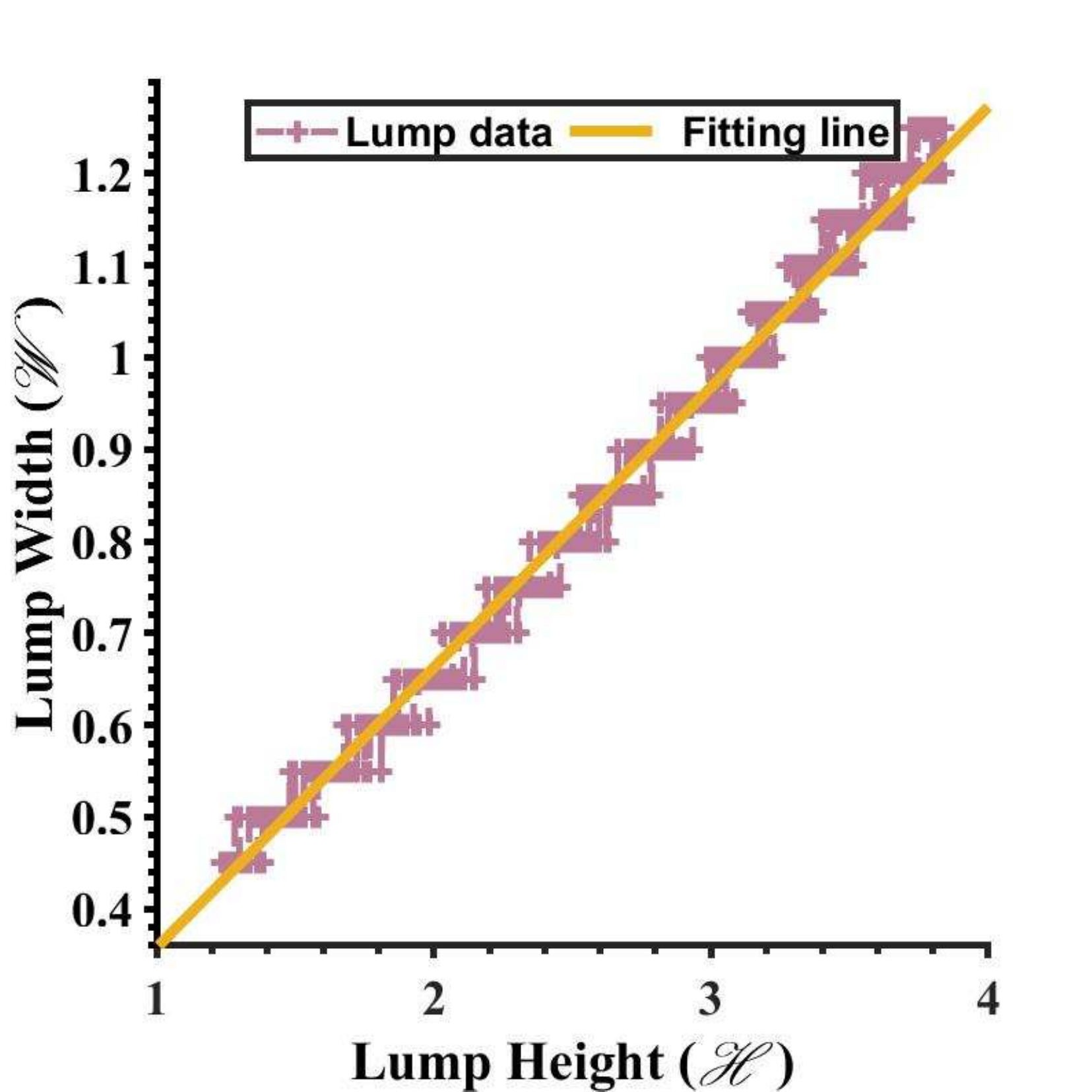}
\end{minipage}
}
\caption{Evolution of a magnetic lump in a damped ferrite film with dimensionless damping constant $s\!=\!0.1$. (a) Comparison picture of lump wave at $t\!=\!0$ and $t\!=\!10$. (b) The variation of lump height $\mathcal{H}$, lump width $\mathcal{W}$ and velocity V. (c) Numerical relationship between the width and height of magnetic lump.}
\end{figure*}
The evolution behavior of the magnetic lump in the ideal ferrite is quite simple and imaginable.
Each lump maintains its shape while it travels at a constant speed.
However, in most of real ferromagnetic materials, we have to take the Gilbert-damping into account .
For instance, the dimensionless damping constant $s$ ranges from 0.048 to about 0.385 in garnet ferrite films.
Here we are going to study the dynamics of magnetic lump in a damped ferrite film.
The typical ferromagnetic film under consideration is a garnet ferrite film with the dimensionless damping constant $s=0.1$.
For a clearer view of the change in shape of the lump, we define $\mathcal{H}$ and $\mathcal{W}$ as the height and width of the lump, which are the vertical distance between the highest point and the lowest point and the horizontal distance along the propagation direction, respectively.
All of these are summarized in Fig.4.

The propagation of a lump on the garnet ferrite film is presented in Fig.5.
As shown in Fig.5(a), the lump travels forward a visible distance in the damped ferrite.
Beyond that, comparing the profiles of lump between $t\!=\!0$ and $t\!=\!10$, we evidently observe that the lump becomes smaller and narrower.
Fig.5(b) shows the lump height and width exhibit a tendency of exponential decay.
The solid blue line is the exponential fitting curve to $\mathcal{H}(t)$, with the function expression being $\mathcal{H}(t)=A_{0}e^{-st}$.
We confirm the above-mentioned amplitude attenuation law is universal by simulating the motion of lump in ferrites with virous damping factors.
Moreover, a definite relationship between the amplitude and the localization region of solitons is important for the soliton excitations. We analyze different sizes of numerical lumps and mark the width and height of lumps in the phase diagram (see Fig. 5(c)). The results show that for a magnetic lump excitation, its width and height meet a linear relationship within the error range ($\mathcal{W}/\mathcal{H}\sim0.305$). So the lump excitation, upon decay, retains a soliton form.
Therefore, in this system, the Gilbert-damping plays a role of dissipating energy during the motion of magnetic lumps and it is characterized by decreasing the amplitude and width of lump.

The inhomogeneities otherwise referred to as deformities is inevitable in real magnetic materials, and it can be caused by either external fields or the presence of defects, voids and gaps in the material.
It has already been reported that the MS may be deformed by the presence of inhomogeneities, in particular its structure and speed \cite{saravanan2018engineering,saravanan2020perturbed}.
In this present system, the inhomogeneous exchange process is unignorable when the wavelength of lump is comparable to the characteristic exchange length.
\begin{figure}[htbp]
\subfigure[]{
\begin{minipage}[t]{0.45\linewidth}
\centering
\includegraphics[width=4cm]{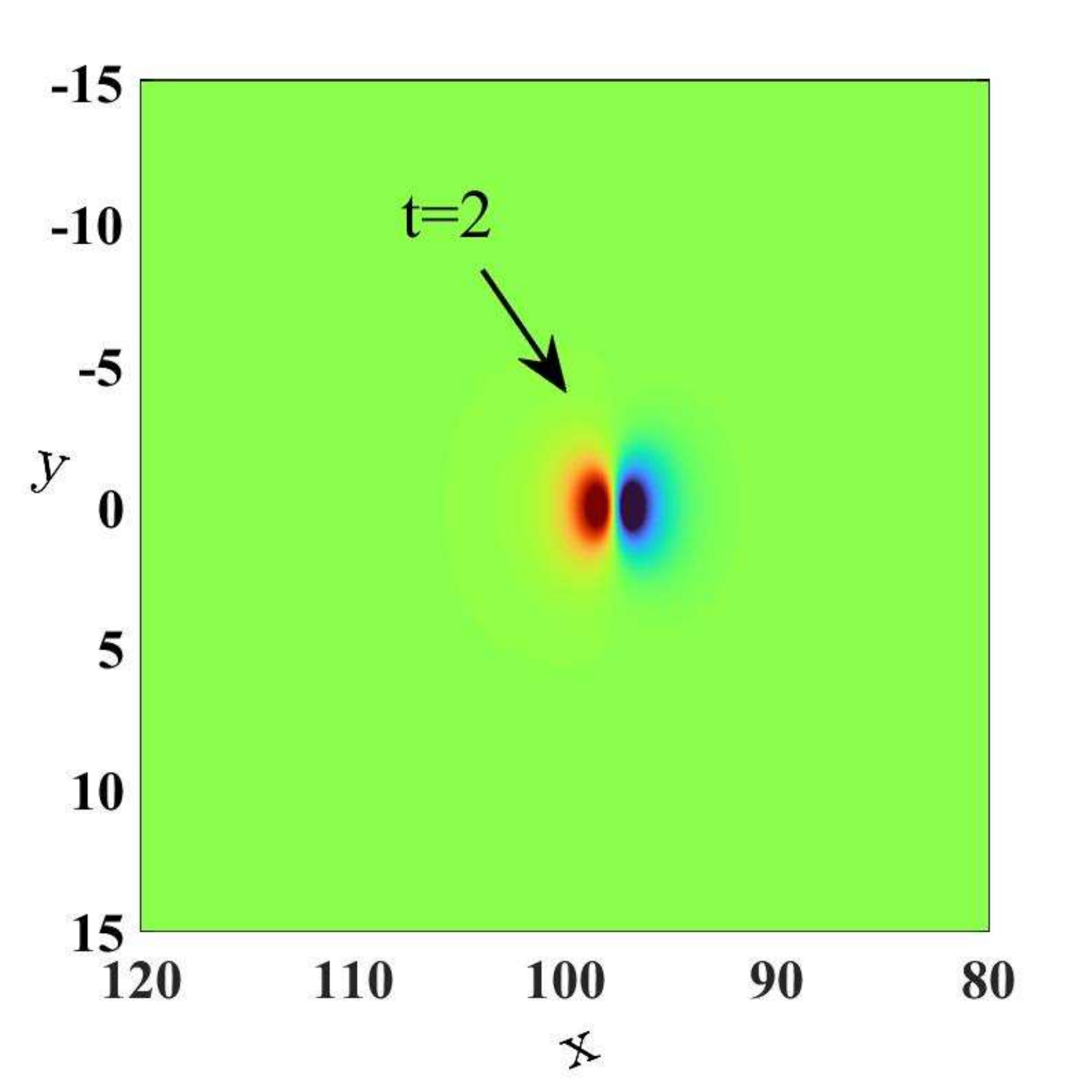}
\end{minipage}
}
\subfigure[]{
\begin{minipage}[t]{0.45\linewidth}
\centering
\includegraphics[width=4cm]{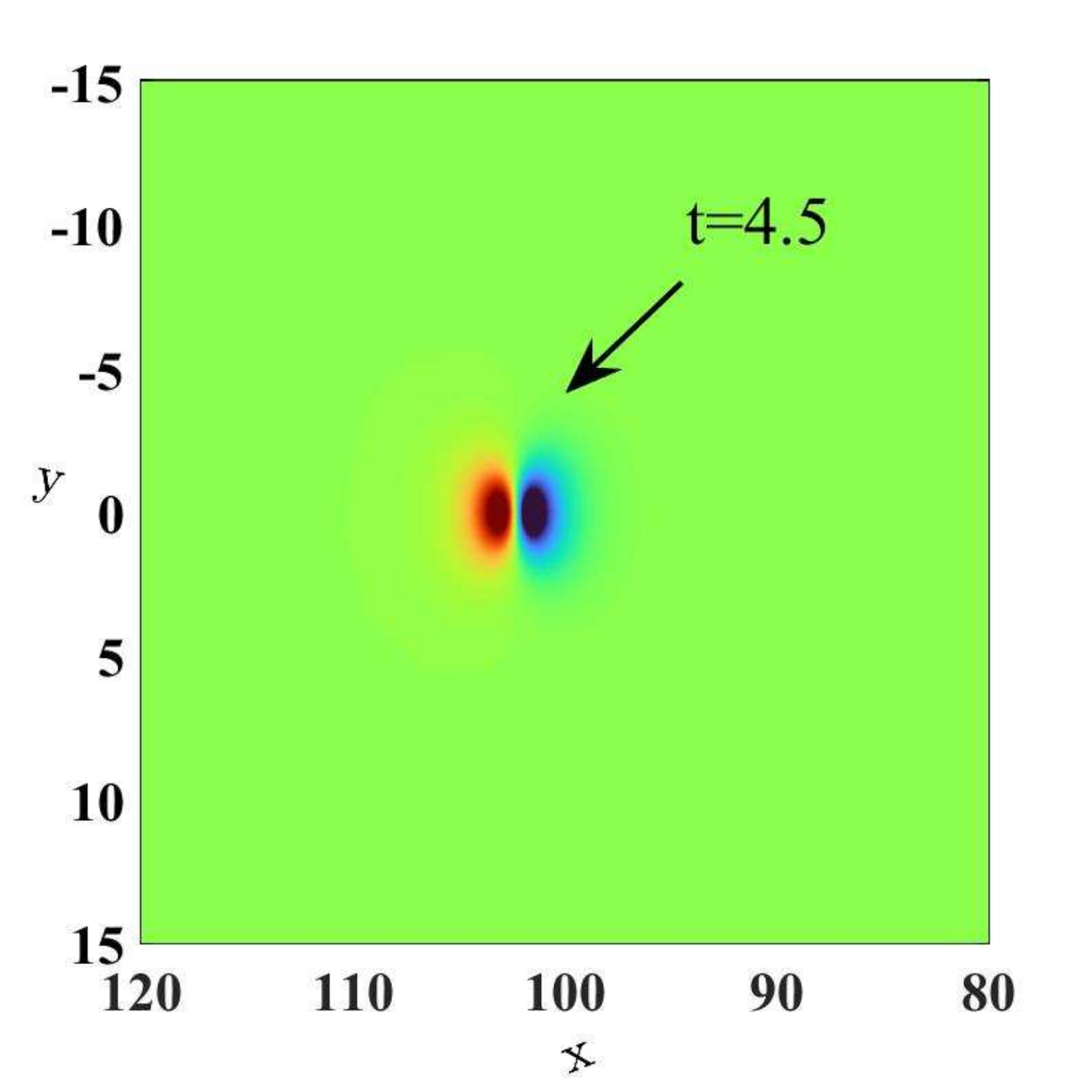}
\end{minipage}
}

\subfigure[]{
\begin{minipage}[t]{0.45\linewidth}
\centering
\includegraphics[width=4cm]{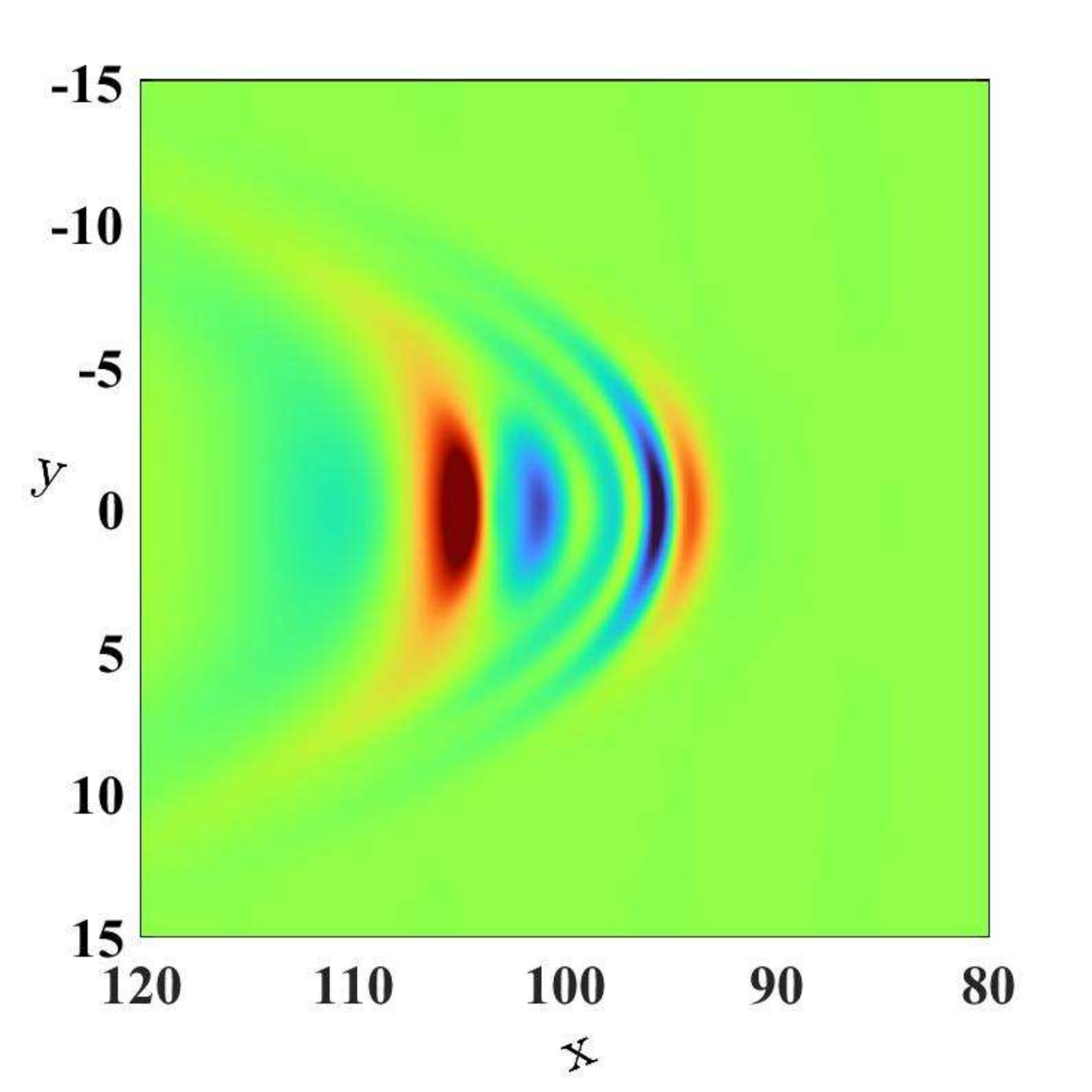}
\end{minipage}
}
\subfigure[]{
\begin{minipage}[t]{0.45\linewidth}
\centering
\includegraphics[width=4cm]{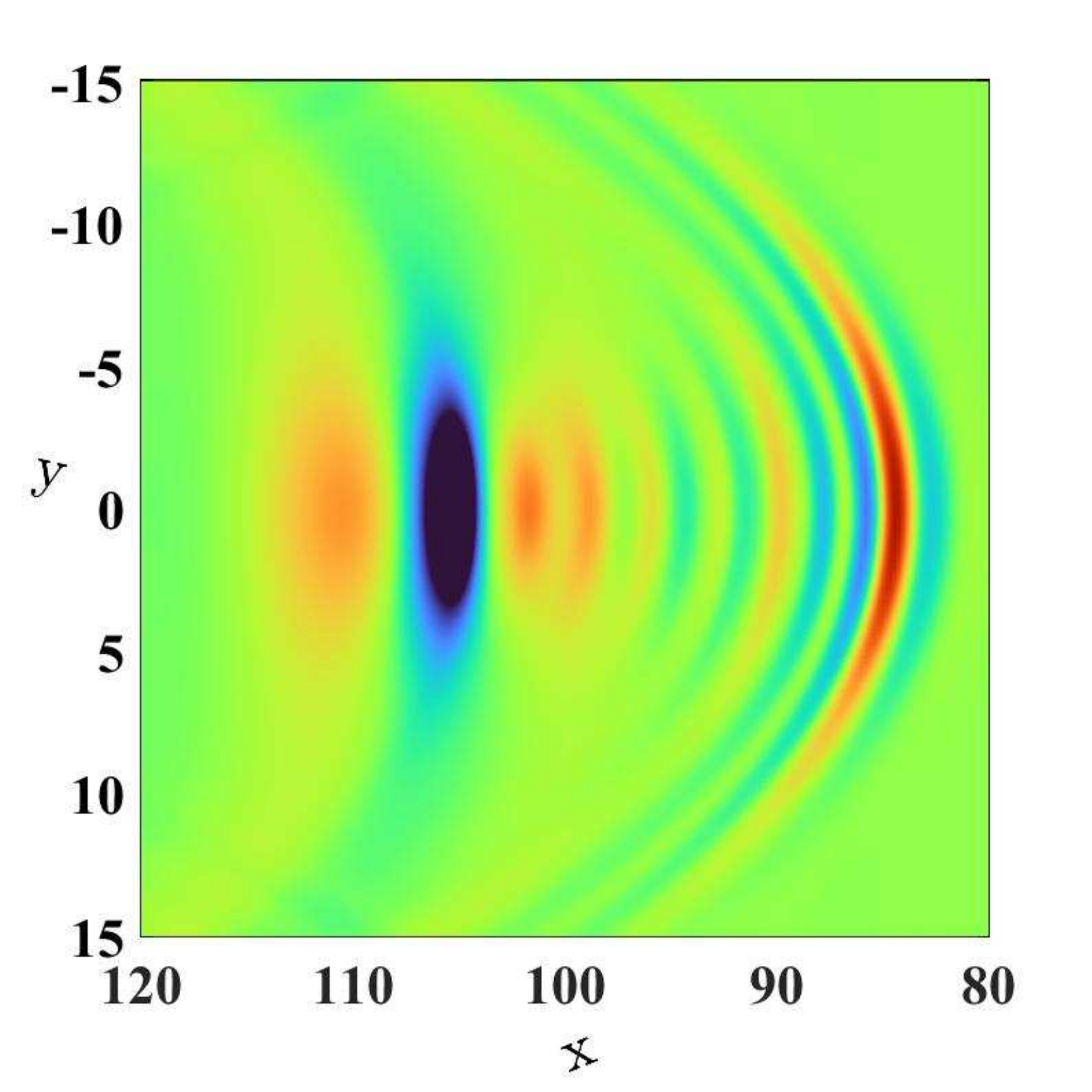}
\end{minipage}
}
\caption{Propagation of lump with and without the inhomogeneous interaction, respectively.}
\end{figure}

We now move to study the lump motion in the presence of inhomogeneous exchange effect.
The initial data is the stable magnetic lump shown in Fig.5.
As can be observed from Fig. 6(a) and 6(b), in ferrite without exchange interaction, the lump solution propagates at a constant speed and along the previous path.
We then consider the non-equilibrium dynamics of lump by performing a sudden interaction quench.
The pictures of component $H^{y}$ at dimensionless times $t\!=\!2$ and $t\!=\!4.5$ are shown in Fig. 6(c) and 6(d).
As we see, for a quench from the non-interacting to strong inhomogeneous exchange ferrite film, the lump oscillates rapidly and diffracts along the propagation direction.
A two-dimensional shock wave is generated and propagates forward.
The shock wave front continues to propagate in the negative direction along $x$-axis.
Finally, the energy of lump will be dissipated into numberless tiny waves.
Accordingly, considering that the lump would be destroyed by the inhomogeneous exchange process, one has to consider keeping its wavelength away from the characteristic exchange length in the lump-based microwave applications.
\begin{figure*}[htbp]
\centering
\vspace{3pt}
\subfigtopskip=2pt
\subfigbottomskip=0pt
\subfigcapskip=0pt
\subfigure[]{
\begin{minipage}[t]{1\linewidth}
\centering
\includegraphics[width=13.5cm]{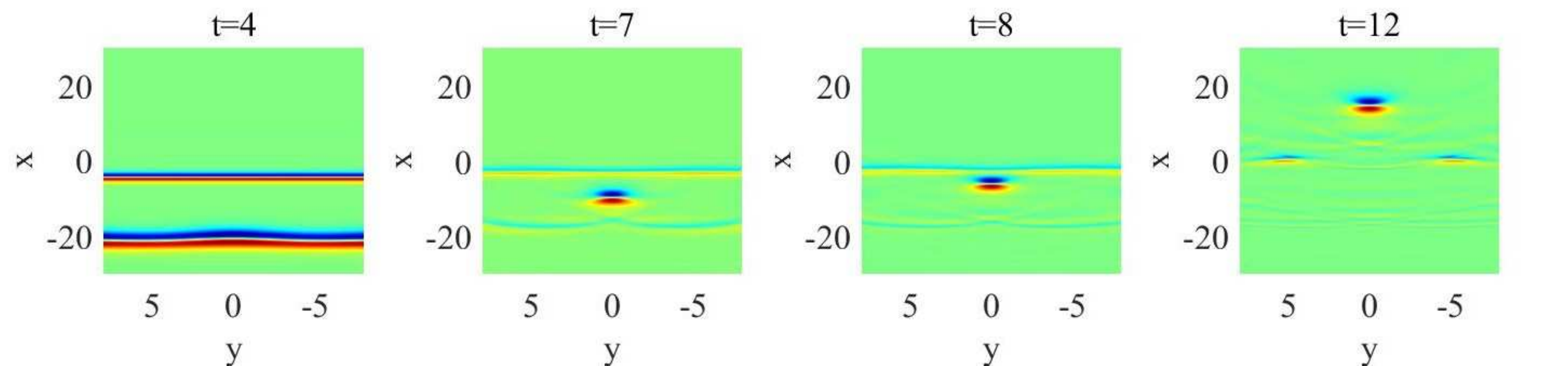}
\end{minipage}
}

\subfigure[]{
\begin{minipage}[t]{1\linewidth}
\centering
\includegraphics[width=13.5cm]{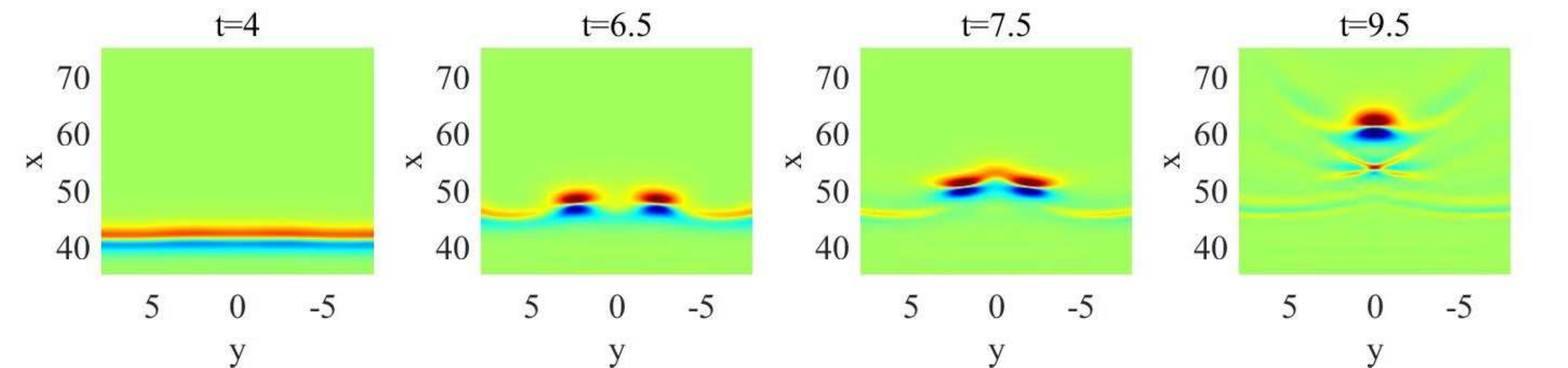}
\end{minipage}
}
\caption{(a) Collision between lump and MS. (b) Mergence of two lumps and the formation of a second-order rogue wave-like structure.}
\end{figure*}

\subsection{Some examples of excitations and interactions}
The evolution pattern given in Fig.2 reveals that the lump moves at a larger velocity than the broken MS in the propagation.
The reason is that the velocity of soliton solution is proportional to the soliton amplitude.
During the formation of the lump, the original MS will be destroyed, and most of the energy is concentrated in some certain centers, which causes the amplitude (and velocity) of the lump to be greater than that of MS.
These lumps with various speeds enable us to explore the interaction between lump and soliton, as well as between two lumps.

A typical example of lump-MS collision is shown in Fig.7(a).
The MS begins to break up around at $t=4$. Subsequently, the splitting lump is going to catch up and collide with the front-MS.
After the collision, the front-MS is destroyed and broken into several lumps with various sizes.
It is remarkable that the lump keep its localized form before and after the collision almost unchanged. This phenomenon implies such two-component lumps are natural results from this nonlinear propagation equations.
Further simulation shows these lump structures could be generated by a MS with random disturbance.
Fig.7(b) depicts a characteristic inelastic collision between two lumps.
We initially generate two adjoining lumps. They are emitted by MS at dimensionless time $t=6.5$.
The merging process can be performed as follows.
From $t=7.5$ to $t=9.5$, two lumps merge simultaneously together and give birth to a new lump whose amplitude is significantly greater than the amplitude of previous lumps.
Obviously there is a weak attraction between two lumps which results in their fusion.
In addition to the fusion of the two lumps, we also observed an extraordinary peak at a specific moment (about $t=9.5$), which looks like a second-order rogue wave.
It appears to be the result of the interaction between the ripples surrounding the two lumps.
After the fusion, the rouge wave-like structure disappears and the dynamics of the output is determined mainly by a single high-amplitude lump.

\section{conclusion}
As a conclusion, the nonlinear propagation of MS in a saturation magnetized ferromagnetic thick film is studied in detail.
In the starting point, we derive the (2+1)-dimensional KMM system that governs the evolution of short MS waves in a saturated ferromagnetic film.
The bilinear form of the KMM system is constructed and the MS solutions are obtained analytically.

After that, numerical simulations are performed to analyse the evolution behaviours of MS.
A significant observation is that the unstable MS can be destroyed by Gaussian perturbation and broken into some stable magnetic lumps.
These lumps exhibit high stability during the propagation.
Furthermore, some examples are given to analyse the collision behaviours between lump and MS, and the interaction between two lumps.
It is found the lump keeps its shape and speed in the collision with MS.
The results confirm that the lump is a stable propagation mode in this system and, more to the point, the velocity of lump can be adjusted by its amplitude.
Their robustness and controllability provide the possibility for future information memory and logic devices.
We also study the propagation of such a lump in ferrites subjected to influence of damping and inhomogeneous exchange effects.
When the Gilbert-damping of ferrite is considered, the lumps undergo the following changes: the amplitude and the speed of lump are decreased, and the width of lump along the propagation direction is getting narrow. It would cause a strong diffraction of the lump if we quench the interaction strength.

We hope our work will invoke follow-up experimental studies of lump-based microwave applications. Additionally, since only one- and two-line-soliton are obtained, the integrability of the (2+1)-dimensional system Kraenkel-Manna-Merle (KMM) remains an open issue.
The existence of the higher-dimensional evolution system as well as the bulk polariton solution is an intriguing avenue for future exploration.

\section*{Acknowledgment}
This work was supported by the National Natural Science Foundation of China under Great Nos. 11835011; 11675146; 11875220;.
\bibliographystyle{unsrt}
\bibliography{references}
\end{document}